\pdfoutput=1
\RequirePackage{ifpdf}
\ifpdf 
\documentclass[pdftex]{sigma}
\else
\documentclass{sigma}
\fi

\theoremstyle{definition}
\newtheorem{cnd}{Condition}

\DeclareMathOperator{\diag}{diag}

\begin{document}

\allowdisplaybreaks

\renewcommand{\PaperNumber}{030}

\FirstPageHeading

\ShortArticleName{Pentagon Relations in Direct Sums and Grassmann Algebras}

\ArticleName{Pentagon Relations in Direct Sums\\
and Grassmann Algebras}

\Author{Igor G.~KOREPANOV and Nurlan M.~SADYKOV}
\AuthorNameForHeading{I.G.~Korepanov and N.M.~Sadykov}

\Address{Moscow State University of Instrument Engineering and Computer Sciences,\\
20 Stromynka Str., Moscow 107996, Russia} 
\Email{\href{mailto:paloff@ya.ru}{paloff@ya.ru}, \href{mailto:nik56-74@mail.ru}{nik56-74@mail.ru}}

\ArticleDates{Received December 19, 2012, in f\/inal form April 05, 2013; Published online April 10, 2013}

\Abstract{We construct vast families of orthogonal operators obeying pentagon relation in a~direct sum of
three $n$-dimensional vector spaces.
As a~consequence, we obtain pentagon relations in Grassmann algebras, making a~far reaching generalization
of exotic Reidemeister torsions.}


\Classification{15A75; 57Q99; 57R56}

\section{Introduction}

\begin{figure}[hb]\centering
\includegraphics[scale=.6]{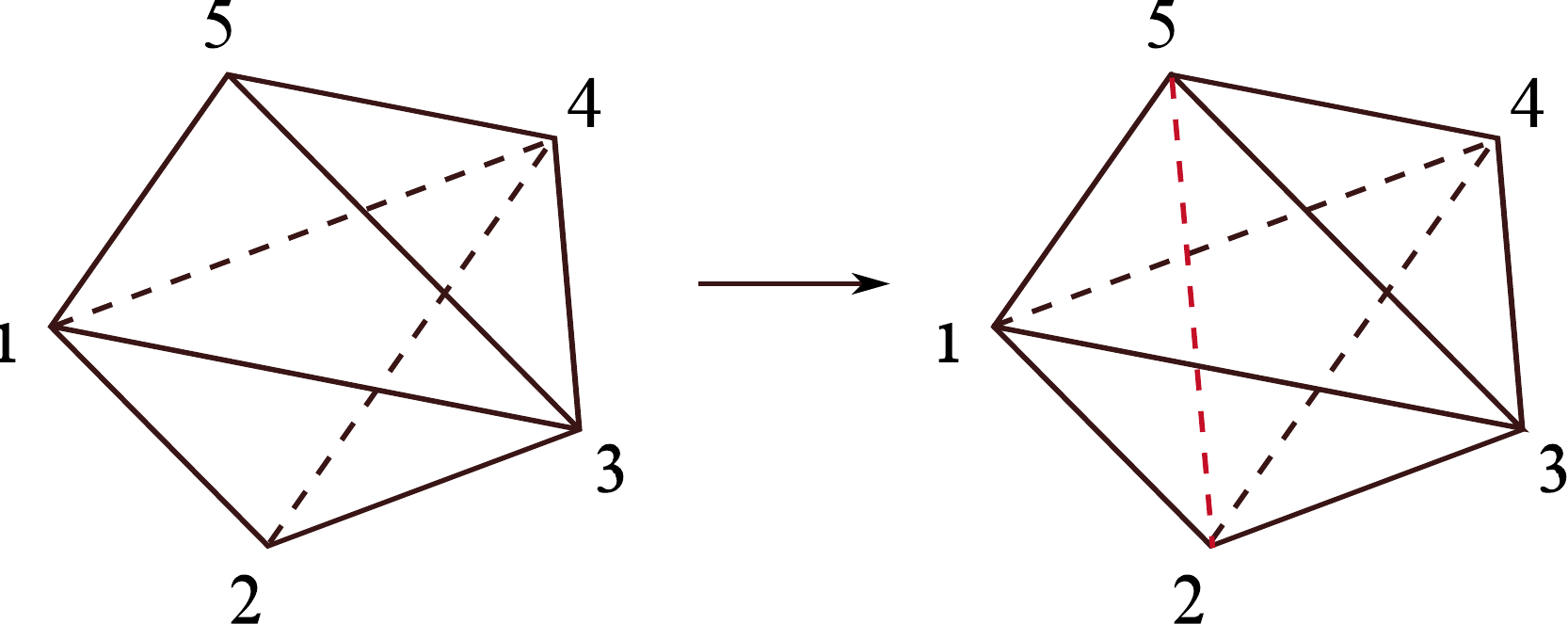}
 \caption{Pachner move $2\to3$.}
\label{f:23}
\end{figure}

\emph{Pachner moves}~\cite{Pachner} are elementary local rebuildings of a~manifold triangulation.
There are $n+1$ types of Pachner moves in dimension~$n$, and a~triangulation of a~piecewise-linear
$n$-manifold can be transformed into another triangulation using a~sequence of these moves; a~pedagogical
introduction in this theory can be found in~\cite{Lickorish}.
In dimension three, the Pachner moves are:
\begin{itemize}\itemsep=0pt
\item move $2\to3$, replacing two adjacent tetrahedra by three tetrahedra occupying the same
place in the manifold, \item inverse move $3\to2$, \item move $1\to4$, decomposing a~tetrahedron into four
tetrahedra with a~new vertex inside it, \item and the inverse move $4\to1$.
\end{itemize}
A move $2\to3$ is depicted in Fig.~\ref{f:23}: the left-hand side consists of tetrahedra $1234$ and~$1345$,
while the right-hand side~-- of tetrahedra $1245$, $2345$ and~$1235$.

We call \emph{pentagon relation} any algebraic formula that corresponds naturally to a~move $2\to3$, in
such way that there is a~hope to move further and develop a~theory that would give some sort of manifold
invariant.
Specif\/ically, this paper deals with pentagon relations in a~direct sum of complex vector spaces, as in
formulas~\eqref{bd} and~\eqref{QP=TSR} below (where matrix entries can be themselves matrices), and in
a~Grassmann algebra, as in formula~\eqref{pGB}.
A subtle point is that a~Grassmann-algebraic relation follows from a~direct-sum relation if the latter is
made of \emph{orthogonal} operators, so the construction of such ``orthogonal'' relations, by means of
ansatz~\eqref{fg},~\eqref{XYsol}, involving some non-commutative algebra, is the central result of this
paper.
\begin{remark}
Our pentagon relations are thus slightly unusual, because usually pentagon relations are written in the
\emph{tensor product} of vector spaces, see for instance~\cite{Kashaev-pentagon} and~\cite{SD}.
Nevertheless, paper~\cite{Kashaev-pentagon} already contains an example of a~relation in direct sum, as we
explain in the end of Section~\ref{s:1d}.
\end{remark}
\begin{remark}
Experience shows that if an interesting formula corresponding to moves $2\leftrightarrow3$ has been found,
then, usually, its counterpart corresponding to moves $1\leftrightarrow4$ can be found too.
\end{remark}

Below,
\begin{itemize}\itemsep=0pt
\item in Section~\ref{s:g}, we explain how a~pentagon relation in a~direct sum follows if
vector spaces are put in correspondence to plane polygons, with the condition that if two of them make
together a~larger polygon, then the two corresponding spaces are added directly,
\item in the warming-up
Section~\ref{s:1d}, we obtain, in our own way, the Kashaev's pentagon relation for \emph{orthogonal}
matrices in the direct sum of \emph{one-dimensional} spaces,
\item in Section~\ref{s:nd}, we present
a~simple general construction of vector spaces needed for pentagon relation, but without the requirement of
orthogonality,
\item in the central Section~\ref{s:o}, we introduce an elegant Euclidian metric in these
vector spaces,
\item in Section~\ref{s:2iso}, we look more closely at the case where the space attached to
each triangle is \emph{two}-dimensional, and the (complex) Euclidian metric is introduced in terms of
\emph{isotropic} vectors that will correspond to fermionic creation-annihilation operators,
\item in
Section~\ref{s:GB}, we brief\/ly recall the fundamentals of Grassmann--Berezin calculus of anticommuting
variables, and write out a~Grassmann-algebraic pentagon relation,
\item in Section~\ref{s:G}, we write out
formulas connecting a~Grassmann--Gaussian exponent taken as a~tetrahedron Grassmann weight, and an
orthogonal matrix,
\item and in Section~\ref{s:exotic}, we show that our construction includes the earlier
introduced Grassmann tetrahedron weights related to exotic Reidemeister torsions.
\end{itemize}

We plan to write one more paper, containing relations corresponding to moves 1--4, construction of a~TQFT,
and calculations for specif\/ic manifolds.

\section{Pentagon relation in a~direct sum of based vector spaces:
\\
generalities}
\label{s:g}

\begin{figure}[t]
\centering
\includegraphics[scale=.65]{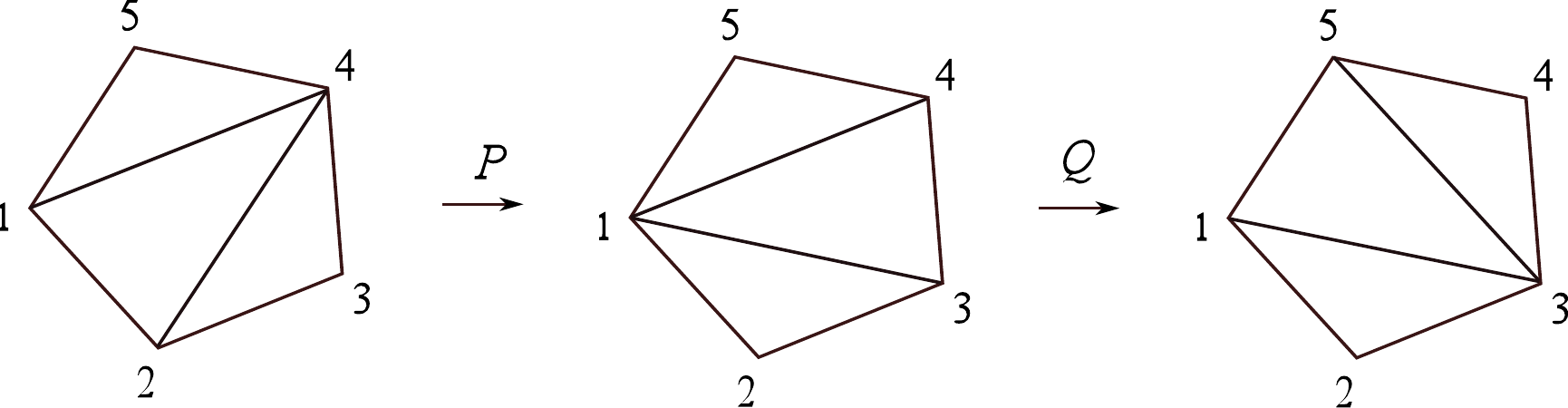}
\caption{Flips in pentagon triangulation corresponding to the l.h.s.\ of Pachner move $2\to 3$.}
\label{f:2}
\bigskip
\centering
\includegraphics[scale=.65]{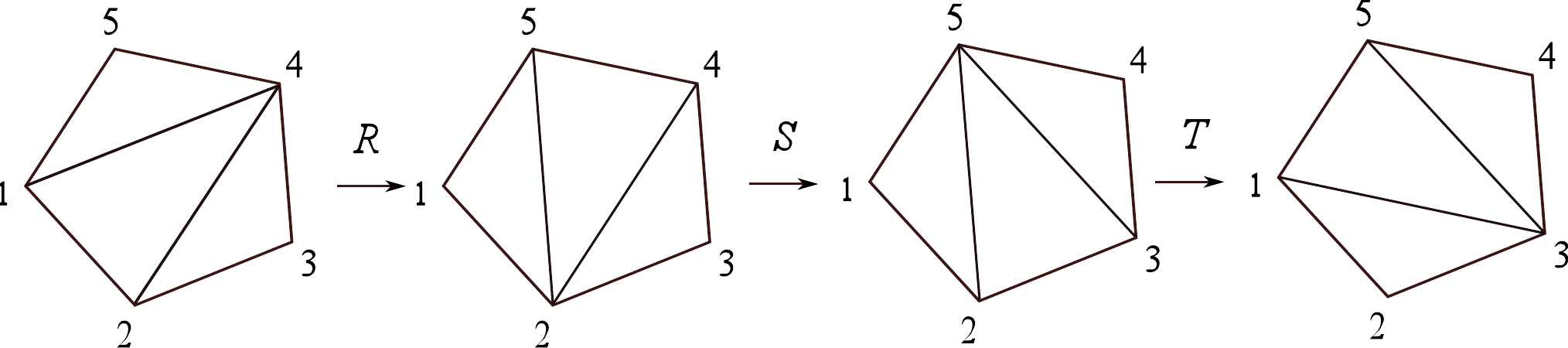}
\caption{Flips in pentagon triangulation corresponding to the r.h.s.\ of Pachner move $2\to 3$.}
\label{f:3}
\end{figure}

We now want to interpret our Fig.~\ref{f:23} as a~f\/lat f\/igure~-- projection of the two sides of the
Pachner move onto the plane.
Thus, for instance, the back surface of two tetrahedra in the left-hand side of Fig.~\ref{f:23} becomes
a~triangulated pentagon~$12345$, made of three triangles $124$, $234$ and~$145$; this is also shown as the
leftmost pentagon in Fig.~\ref{f:2} or~\ref{f:3}.
We put in correspondence to this f\/lat geometric picture the following algebraic objects.
An $n$-dimensional complex vector space will correspond to each triangle with vertices in $\{1,\dots,5\}$,
and we require the following condition.
\begin{cnd}
\label{c:s}
If some triangles make together a~triangulation of a~greater polygon (quadrilateral or the whole pentagon),
then the direct sum of the corresponding spaces depends only on this polygon, i.e., is the same for other
triangulation(s).
\end{cnd}

Thus, all convex quadrilaterals, as well as the pentagon~$12345$, acquire their vector spaces as well.
Taking some liberty, we denote these spaces the same way as polygons, and write Condition~\ref{c:s} as
\begin{gather}
\label{lb}
1234=124\oplus234=123\oplus134,
\qquad
12345=124\oplus234\oplus145=\cdots,
\qquad
\text{etc.}
\end{gather}

We further assume that our spaces corresponding to \emph{triangles} are \emph{based}~-- equipped with
chosen bases.
The bases will be written as columns made of basis vectors, for instance,
\begin{gather*}
\mathbf f_{124}=
\begin{pmatrix}
\mathbf f_{124}^{(1)}
\\
\vdots
\\
\mathbf f_{124}^{(n)}
\end{pmatrix}
.
\end{gather*}
For a~greater polygon, any decomposition like those in~\eqref{lb} provides a~basis in the form of the
ordered union of bases in triangles, written as, for instance, $
\begin{pmatrix}
\mathbf f_{124}
\\
\mathbf f_{234}
\end{pmatrix}
$ if we are considering the decomposition $1234=124\oplus234$.
\begin{remark}
The specif\/ic order of bases in this union can be f\/ixed in any convenient way.
In the present paper, it always complies with the following rule: $\mathbf f_{ijk}$ goes before $\mathbf
f_{i'j'k'}$ if $i+j+k<i'+j'+k'$.
Actually, this order arose when we were doing calculations in what we were seeing as the most natural way.
We never meet here a~situation where $i+j+k=i'+j'+k'$.
\end{remark}
\begin{remark}
The vertices of a~triangle~$ijk$ always go in this paper in the increasing order: $i<j<k$.
Note that this order induces a~\emph{consistent orientation} of all triangles in all pentagons in
Figs.~\ref{f:2} and~\ref{f:3}.
As we are dealing here only with f\/ive vertices and ten triangles, the orientation issues do not bring
about any problems; we will, however, comment on their importance in a~specif\/ic situation below in
Remark~\ref{r:o}.
\end{remark}

The l.h.s.\ of Fig.~\ref{f:23} can be seen as the tetrahedra $1234$ and~$1345$ glued to the back
surface~$12345$.
In our f\/lat picture, each tetrahedron corresponds to a~\emph{flip}, or two-dimensional Pachner move, in
the triangulation of~$12345$; the sequence of these two f\/lips is depicted in Fig.~\ref{f:2}.

We denote these f\/lips $P$ and~$Q$, and use the same notations for their corresponding matrices def\/ined
as the following change-of-basis matrices in~$12345$:
\begin{gather}
\begin{pmatrix}
\mathbf f_{124}
\\
\mathbf f_{234}
\\
\mathbf f_{145}
\end{pmatrix}
\stackrel{\textstyle P}{\longmapsto}
\begin{pmatrix}
\mathbf f_{123}
\\
\mathbf f_{134}
\\
\mathbf f_{145}
\end{pmatrix}
\stackrel{\textstyle Q}{\longmapsto}
\begin{pmatrix}
\mathbf f_{123}
\\
\mathbf f_{135}
\\
\mathbf f_{345}
\end{pmatrix}
.\label{PQ}
\end{gather}
Sometimes it will be convenient for us to write this also in the following ``matrix${}\times{}$basis'' form:
\begin{gather*}
P
\begin{pmatrix}
\mathbf f_{124}
\\
\mathbf f_{234}
\\
\mathbf f_{145}
\end{pmatrix}
=
\begin{pmatrix}
\mathbf f_{123}
\\
\mathbf f_{134}
\\
\mathbf f_{145}
\end{pmatrix}
,
\qquad
Q
\begin{pmatrix}
\mathbf f_{123}
\\
\mathbf f_{134}
\\
\mathbf f_{145}
\end{pmatrix}
=
\begin{pmatrix}
\mathbf f_{123}
\\
\mathbf f_{135}
\\
\mathbf f_{345}
\end{pmatrix}
.
\end{gather*}

Similarly, the r.h.s.\ of Fig.~\ref{f:23} can be seen as the tetrahedra $1245$, $2345$, and~$1235$ glued to
the back surface~$12345$.
In our f\/lat picture, this corresponds to the sequence of three f\/lips in the triangulation of~$12345$
depicted in Fig.~\ref{f:3}.
We denote these f\/lips $R$, $S$ and~$T$, and their corresponding matrices are determined by the following
actions on bases of~$12345$:
\begin{gather}
\label{RST}
\begin{pmatrix}
\mathbf f_{124}
\\
\mathbf f_{234}
\\
\mathbf f_{145}
\end{pmatrix}
\stackrel{\textstyle R}{\longmapsto}
\begin{pmatrix}
\mathbf f_{125}
\\
\mathbf f_{234}
\\
\mathbf f_{245}
\end{pmatrix}
\stackrel{\textstyle S}{\longmapsto}
\begin{pmatrix}
\mathbf f_{125}
\\
\mathbf f_{235}
\\
\mathbf f_{345}
\end{pmatrix}
\stackrel{\textstyle T}{\longmapsto}
\begin{pmatrix}
\mathbf f_{123}
\\
\mathbf f_{135}
\\
\mathbf f_{345}
\end{pmatrix}.
\end{gather}
\begin{theorem}
Matrices $P,\ldots,T$ have the following forms:
\begin{gather}
P=
\begin{pmatrix}*&*&0\\ {*}&*&0\\0&0&1\end{pmatrix}
,
\qquad
Q=
\begin{pmatrix}1&0&0\\0&*&*\\0&*&*\end{pmatrix},
\nonumber\\
R=
\begin{pmatrix}*&0&*\\ 0&1&0\\ {*}&0&*\end{pmatrix}
,
\qquad
S=
\begin{pmatrix}1&0&0\\0&*&*\\0&*&*\end{pmatrix}
,
\qquad
T=
\begin{pmatrix}*&*&0\\ {*}&*&0\\0&0&1\end{pmatrix}
,
\label{bd}
\end{gather}
where each zero, one or asterisk is an $n\times n$ matrix block, and the following \emph{pentagon relation
in the direct sum} holds:
\begin{gather}
\label{QP=TSR}
QP=TSR.
\end{gather}
\end{theorem}
\begin{proof}
The form~\eqref{bd} of matrices is clear from~\eqref{PQ} and~\eqref{RST}, as well as the
relation~\eqref{QP=TSR} that follows from the fact that both leftmost and rightmost bases in~\eqref{PQ}
and~\eqref{RST} are the same.
\end{proof}
\begin{remark}
\label{r:i}
We can now forget about the space~$12345$ and its subspaces and consider matrices $P,\dots,T$ simply as
linear operators acting in the $3n$-dimensional space of usual column vectors~-- the direct sum of three
such $n$-dimensional spaces.
Each of the matrices $P,\ldots,T$ acts nontrivially in the direct sum of \emph{two} such spaces, and can be
regarded, in the spirit of similar relations in mathematical physics, as a~$2n\times2n$ matrix,
identif\/ied when necessary with its direct sum with an identity matrix.
\end{remark}
\begin{remark}
Equation~\eqref{QP=TSR} for matrices~\eqref{bd} can be regarded as a~simplif\/ied version of ``dynamical
Yang--Baxter equation in a~direct sum'', introduced by one of the authors~\cite{K94} in~1994.
The latter equation appears at each step of evolution of an integrable dynamical system in discrete time,
see~\cite[formulas~(3),~(4) and~(10)]{K94}, and proved its usefulness and fundamental character both in
dynamical systems and constructing solutions to Zamolodchikov tetrahedron equation, enough to mention
papers~\cite{Kashaev95,FTE,KuSe,S07,S06}.
\end{remark}

Thus, if we construct $n$-dimensional subspaces of a~$3n$-dimensional vector space as described in the
beginning of this section, including Condition~\ref{c:s}, we get matrices satisfying the pentagon relation
in the direct sum.
This turns out to be, in itself, quite easy; interesting complications come forward, as we will see, when
there is additional requirement of orthogonality.

\section{Kashaev's pentagon relation for orthogonal matrices
\\
in a~direct sum of one-dimensional spaces}
\label{s:1d}

In this section, we take the following space~$V$ as the space~$12345$ introduced in Section~\ref{s:g}: by
def\/inition, $V$ is the three-dimensional complex vector space of row vectors of length~5
\begin{gather}
\label{f}
\mathbf f=
\begin{pmatrix}f_1&f_2&f_3&f_4&f_5\end{pmatrix}
\end{gather}
subject to the conditions
\begin{gather}
f_1+f_2+f_3+f_4+f_5=0,
\qquad
\zeta_1f_1+\zeta_2f_2+\zeta_3f_3+\zeta_4f_4+\zeta_5f_5=0,
\label{fcond}
\end{gather}
where $\zeta_i$ are some constants, all pairwise dif\/ferent.
We will be using the following notation for their dif\/ferences:
\begin{gather}
\label{zij}
\zeta_{ij}\stackrel{\rm def}{=}\zeta_i-\zeta_j.
\end{gather}
\begin{remark}
Our motivation for introducing the vectors~\eqref{f} with conditions~\eqref{fcond} here, and their
generalizations below, was simply that we have already used such objects in our earlier works, although in
a~dif\/ferent context.
It is enough to mention the paper~\cite{11S} and formulas (8), (9) and the unnumbered one, next after~(33),
therein.
\end{remark}

For $1\le i<j<k\le5$, choose vectors~$\mathbf f_{ijk}$ with only $i$-th, $j$-th and~$k$-th nonzero
components.
This means that, for instance,
\begin{gather*}
\mathbf f_{135}=\mathrm{const}_{135}
\begin{pmatrix}\zeta_{35}&0&\zeta_{51}&0&\zeta_{13}\end{pmatrix}
.
\end{gather*}
Such a~vector forms itself a~basis in the one-dimensional space~$ijk$, corresponding to the triangle~$ijk$.
At this stage, the nonzero constants $\mathrm{const}_{ijk}$ can be arbitrary.

Now we def\/ine $3\times3$ matrices $P$, $Q$, $R$, $S$ and~$T$ by the conditions~\eqref{PQ}
and~\eqref{RST}, and get a~solution to~\eqref{QP=TSR} nicely parametrized by numbers~$\zeta_i$.
In particular, the matrix corresponding to a~tetrahedron depends only on the four $\zeta_i$'s at its
vertices.

Next, we want to make \emph{orthogonal} operators from our matrices $P,\ldots,T$.
This is done in two steps.
First, we introduce a~scalar product between vectors~\eqref{f},~\eqref{fcond}, and second, we choose
constants $\mathrm{const}_{ijk}$, in such way that the following conditions hold.
\begin{cnd}
\label{c:1o}
Scalar product between vectors~\eqref{f},~\eqref{fcond} must be such that all three vectors in any column
in formulas~\eqref{PQ} and~\eqref{RST} become pairwise orthogonal.
\end{cnd}

\begin{cnd}
\label{c:1u}
The normalizing constants $\mathrm{const}_{ijk}$ are such that all $\mathbf f_{ijk}$ become unit vectors.
\end{cnd}

Condition~\ref{c:1o} applies to vectors corresponding to triangles with disjoint interiors, i.e., any two
triangles within any chosen pentagon in Figs.~\ref{f:2} and~\ref{f:3}; it can be checked that Condition~\ref{c:1o} for all such triangles follows from
the f\/ive independent conditions
\begin{gather}
\label{ort}
123\perp345,
\qquad
234\perp145,
\qquad
345\perp125,
\qquad
145\perp123,
\qquad
125\perp234.
\end{gather}
As a~symmetric $3\times3$ matrix determining the (complex) Euclidian metric in our three-dimensional
space~$V$ has six independent entries, it is determined by the f\/ive linear conditions~\eqref{ort}
uniquely up to a~scalar factor.
Then, Condition~\ref{c:1u} determines uniquely all normalizers $\mathrm{const}_{ijk}$.

We do not present here explicit formulas for the scalar product and normalizing constants, as they are
written out below (formulas~\eqref{fg} and~\eqref{XYsol}) in a~more general case.
The resulting pentagon relation for orthogonal matrices is:
\begin{gather}
\begin{pmatrix}
1&0&0
\\
0&\cos\varphi_{1345}&\sin\varphi_{1345}
\\
0&-\sin\varphi_{1345}&\cos\varphi_{1345}
\end{pmatrix}
\begin{pmatrix}
\cos\varphi_{1234}&-\sin\varphi_{1234}&0
\\
\sin\varphi_{1234}&\cos\varphi_{1234}&0
\\
0&0&1
\end{pmatrix}
\label{o}
\\
=
\begin{pmatrix}
\cos\varphi_{1235}&-\sin\varphi_{1235}&0
\\
\sin\varphi_{1235}&\cos\varphi_{1235}&0
\\
0&0&1
\end{pmatrix}
\begin{pmatrix}
1&0&0
\\
0&\cos\varphi_{2345}&\sin\varphi_{2345}
\\
0&-\sin\varphi_{2345}&\cos\varphi_{2345}
\end{pmatrix}
\begin{pmatrix}
\cos\varphi_{1245}&0&\sin\varphi_{1245}
\\
0&1&0
\\
-\sin\varphi_{1245}&0&\cos\varphi_{1245}
\end{pmatrix}\!,
\nonumber
\end{gather}
where the cosines and sines are the following square roots of cross-ratios:
\begin{gather*}
\cos\varphi_{ijkl}=\sqrt{\frac{\zeta_{il}\zeta_{jk}}{\zeta_{ik}\zeta_{jl}}},
\qquad
\sin\varphi_{ijkl}=\sqrt{\frac{\zeta_{ij}\zeta_{kl}}{\zeta_{ik}\zeta_{jl}}},
\end{gather*}
and the exact sense of square roots may be, for instance, as follows: assume
\begin{gather*}
\zeta_1>\zeta_2>\zeta_3>\zeta_4>\zeta_5
\end{gather*}
and take positive square roots.
Computer algebra system Maxima\footnote{Maxima, a~computer algebra system, \url{http://maxima.sourceforge.net}.} 
was of great assistance in deriving and checking
relation~\eqref{o}.
Note again the nice parametrization: each of the matrices depends only on the $\zeta_i$'s in the vertices
of the corresponding tetrahedron.
We emphasize this fact because this will be our goal in the more complicated situation of
Section~\ref{s:o}, where our $\zeta_i$'s will be $n\times n$ matrices.

A relation equivalent to~\eqref{o} can be found in Kashaev's paper~\cite[Example~3]{Kashaev-pentagon}.

\section[A construction of based $n$-dimensional spaces yielding pentagon relation]{A construction of based $\boldsymbol{n}$-dimensional spaces  \\
yielding pentagon relation}
\label{s:nd}

Let now matrix entries in~\eqref{bd}~-- zeros, ones and asterisks~-- be themselves matrices of size
$n\times n$.
As long as we are not concerned about orthogonality, the construction of a~pentagon relation goes
practically the same way as in the beginning of Section~\ref{s:1d}.

To be more exact, we now take for our space $V=12345$ the $3n$-dimensional vector space of row vectors of
length~$5n$ subject to $2n$ linear dependencies.
The row vectors will be written as
\begin{gather}
\label{sf}
\begin{pmatrix}
\mathsf f_1&\mathsf f_2&\mathsf f_3&\mathsf f_4&\mathsf f_5
\end{pmatrix}
,
\end{gather}
where each sans serif letter $\mathsf f_i$ denotes, in its turn, a~row vector of $n$ components, and the
linear dependencies are
\begin{gather*}
\mathsf f_1+\mathsf f_2+\mathsf f_3+\mathsf f_4+\mathsf f_5=0,
\qquad
\mathsf f_1\zeta_1+\mathsf f_2\zeta_2+\mathsf f_3\zeta_3+\mathsf f_4\zeta_4+\mathsf f_5\zeta_5=0,
\end{gather*}
where $\zeta_i$ are $n\times n$ matrices, with the condition that all their dif\/ferences
\begin{gather*}
\zeta_{ij}=\zeta_i-\zeta_j
\end{gather*}
are invertible.

We also introduce a~notation for an $n\times5n$ matrix made of $n$ vectors of the form~\eqref{sf}:
\begin{gather*}
\mathbf f=
\begin{pmatrix}
\mathsf f_1^{(1)}&\mathsf f_2^{(1)}&\mathsf f_3^{(1)}&\mathsf f_4^{(1)}&\mathsf f_5^{(1)}
\\
\vdots&\vdots&\vdots&\vdots&\vdots
\\
\mathsf f_1^{(n)}&\mathsf f_2^{(n)}&\mathsf f_3^{(n)}&\mathsf f_4^{(n)}&\mathsf f_5^{(n)}
\end{pmatrix}
=
\begin{pmatrix}
f_1&f_2&f_3&f_4&f_5
\end{pmatrix}
,
\end{gather*}
where each $f_i=
\begin{pmatrix}
\mathsf f_i^{(1)}
\\
\vdots
\\
\mathsf f_i^{(n)}
\end{pmatrix}
$ is, of course, an $n\times n$ matrix.

In analogy with Section~\ref{s:1d}, we introduce block matrices~$\mathbf f_{ijk}$ with only $i$-th, $j$-th
and~$k$-th nonvanishing $n\times n$ matrix components.
This means that, for instance,
\begin{gather}
\label{f135}
\mathbf f_{135}=
\begin{pmatrix}
f_1&0&f_1\zeta_{51}\zeta_{35}^{-1}&0&f_1\zeta_{13}\zeta_{35}^{-1}
\end{pmatrix}
.
\end{gather}
We also assume the $n$ rows of such matrices to be \emph{linearly independent}; they form then a~basis in
a~corresponding space~$ijk$, for instance, space~$135$ in the case of matrix~\eqref{f135}.
Then we can def\/ine $3n\times3n$ matrices $P$, $Q$, $R$, $S$ and~$T$ by the conditions~\eqref{PQ}
and~\eqref{RST}, and these conditions again ensure the pentagon relation~\eqref{QP=TSR}.

Let us write out explicit formulas for the matrix~$P$ in the case where the bases in relevant spaces are
chosen as
\begin{gather}
\mathbf f_{124}=
\begin{pmatrix}
1&\zeta_{14}\zeta_{42}^{-1}&0&\zeta_{12}\zeta_{24}^{-1}
\end{pmatrix},
\qquad
\mathbf f_{234}=
\begin{pmatrix}
0&1&\zeta_{24}\zeta_{43}^{-1}&\zeta_{23}\zeta_{34}^{-1}
\end{pmatrix},
\nonumber\\
\mathbf f_{123}=
\begin{pmatrix}
1&\zeta_{13}\zeta_{32}^{-1}&\zeta_{12}\zeta_{23}^{-1}&0
\end{pmatrix}, \qquad \mathbf f_{134}=
\begin{pmatrix}
1&0&\zeta_{14}\zeta_{43}^{-1}&\zeta_{13}\zeta_{34}^{-1}
\end{pmatrix}
\label{f124}
\end{gather}
(compare this to formula~\eqref{f135} with $f_1$ set to identity matrix).
Here we have left out the irrelevant (and zero) f\/ifth matrix component of the vectors; recall also that,
according to Remark~\ref{r:i}, we identify matrices like~$P$ with their direct sums with identity matrices.
A direct calculation shows that the condition
\begin{gather*}
P
\begin{pmatrix}
\mathbf f_{124}
\\
\mathbf f_{234}
\end{pmatrix}
=
\begin{pmatrix}
\mathbf f_{123}
\\
\mathbf f_{134}
\end{pmatrix}
\end{gather*}
leads to the following explicit expression:
\begin{gather}
\label{P}
P=
\begin{pmatrix}
1&\zeta_{12}\zeta_{23}^{-1}\zeta_{34}\zeta_{42}^{-1}
\vspace{1mm}\\
1&-\zeta_{14}\zeta_{42}^{-1}
\end{pmatrix}
.
\end{gather}

The problem of introducing a~Euclidian metric making our operators $P,\dots,T$ orthogonal is not so easy as
in Section~\ref{s:1d}, and deserves a~separate section.

\section{Making operators orthogonal}
\label{s:o}

We begin with analyzing how to introduce complex Euclidian metric in the space~$1234$ where operator~$P$
acts nontrivially, aiming at making $P$ orthogonal; the same reasonings will apply to the other operators
$Q,\dots,T$, after obvious changes of indices.
Space~$1234$ consists of $4n$-row vectors
\begin{gather}
\label{4f}
\begin{pmatrix}
\mathsf f_1&\mathsf f_2&\mathsf f_3&\mathsf f_4
\end{pmatrix}
\end{gather}
with linear relations
\begin{gather}
\label{4c}
\mathsf f_1+\mathsf f_2+\mathsf f_3+\mathsf f_4=0
\qquad
\text{and}
\qquad
\mathsf f_1\zeta_1+\mathsf f_2\zeta_2+\mathsf f_3\zeta_3+\mathsf f_4\zeta_4=0.
\end{gather}
There are two conditions that can be considered separately, like we did with Conditions~\ref{c:1o}
and~\ref{c:1u} in Section~\ref{s:1d}.
\begin{cnd}
\label{i}
The decompositions
\begin{gather*}
1234=124\oplus234=123\oplus134
\end{gather*}
must be orthogonal:
\begin{gather}
\label{perp}
124\perp234,
\qquad
123\perp134.
\end{gather}
\end{cnd}
\begin{cnd}
\label{ii}
The bases within each of the spaces $124$, $234$, $123$ and~$134$ must be orthonormal.
\end{cnd}

Condition~\ref{i} means that, for the matrices def\/ined by formulas~\eqref{f124}, the rows
of matrix~$\mathbf f_{124}$ must be orthogonal to the rows of matrix~$\mathbf f_{234}$, and the rows of
matrix~$\mathbf f_{123}$ must be orthogonal to the rows of matrix~$\mathbf f_{134}$.

We are searching for \emph{some} complex Euclidian scalar product between vectors~\eqref{4f},~\eqref{4c},
and with as good properties as possible.
It turns out to be a~good idea to take this scalar product in the following form:
\begin{gather}
\label{fg}
\Bigl(
\begin{pmatrix}
\mathsf f_1&\mathsf f_2&\mathsf f_3&\mathsf f_4
\end{pmatrix}
, \
\begin{pmatrix}
\mathsf g_1&\mathsf g_2&\mathsf g_3&\mathsf g_4
\end{pmatrix}
\Bigr)=\mathsf f_1X\mathsf g_2^{\mathrm T}+\mathsf g_1X\mathsf f_2^{\mathrm T}
+\mathsf f_3Y\mathsf g_4^{\mathrm T}+\mathsf g_3Y\mathsf f_4^{\mathrm T},
\end{gather}
where $X$ and~$Y$ are $n\times n$ matrices to be found.
The orthogonality conditions are re-written as follows:
\begin{gather}
\label{XY}
X+\zeta_{24}\zeta_{43}^{-1}Y\zeta_{24}^{-\mathrm T}\zeta_{12}^{\mathrm T}=0,
\qquad
X\zeta_{32}^{-\mathrm T}\zeta_{13}^{\mathrm T}+\zeta_{12}\zeta_{23}^{-1}Y\zeta_{34}^{-\mathrm T}\zeta_{13}
^{\mathrm T}=0,
\end{gather}
where we use the notation $\zeta_{ij}^{-\mathrm T}\stackrel{\rm def}{=}\big(\zeta_{ij}^{\mathrm T}\big)^{-1}$.
Eliminating~$Y$, we get the equation for~$X$:
\begin{gather}
\label{X}
X=\zeta_{12}\zeta_{23}^{-1}\zeta_{43}\zeta_{24}^{-1}X\zeta_{12}^{-\mathrm T}\zeta_{24}^{\mathrm T}\zeta_{34}
^{-\mathrm T}\zeta_{32}^{\mathrm T}.
\end{gather}

Next good idea is to take all $\zeta_i$, and thus $\zeta_{ij}$, \emph{symmetric}:
\begin{gather*}
\zeta_{ij}=\zeta_{ij}^{\mathrm T}.
\end{gather*}
Then equations~\eqref{XY} admit the following simple solution:
\begin{gather}
\label{XYsol}
X=\zeta_{12},
\qquad
Y=\zeta_{34}.
\end{gather}
\begin{remark}
Actually, equation~\eqref{X}, for symmetric $\zeta_i$'s with invertible dif\/ferences, can be solved
completely, because it reduces to the form $XAB=BAX$.
Our specif\/ic solution~\eqref{XYsol} seems, however, to be the best, because it also satisf\/ies
Condition~\ref{iii} below.
\end{remark}

Recall that the rows of the f\/irst matrix in~\eqref{f124} form a~basis in the space~$124$, whose vectors we will write,
for a~moment, like
\begin{gather}
\label{124}
\begin{pmatrix}
\mathsf f_1&\mathsf f_2&\mathsf f_4
\end{pmatrix}
,
\qquad
\mathsf f_2=\mathsf f_1\zeta_{14}\zeta_{42}^{-1},
\qquad
\mathsf f_4=\mathsf f_1\zeta_{12}\zeta_{24}^{-1}.
\end{gather}
\begin{theorem}
\label{th:124}
Scalar product~\eqref{fg} for vectors~\eqref{124}, given our solution~\eqref{XYsol}, can be written in
three equivalent forms:
\begin{gather}
\big(
\begin{pmatrix}
\mathsf f_1&\mathsf f_2&\mathsf f_4
\end{pmatrix}
,
\begin{pmatrix}
\mathsf g_1&\mathsf g_2&\mathsf g_4
\end{pmatrix}
\big)\nonumber\\
\qquad{}=\mathsf f_1\zeta_{12}\mathsf g_2^{\mathrm T}+\mathsf g_1\zeta_{12}\mathsf f_2^{\mathrm T}
=\mathsf f_2\zeta_{24}\mathsf g_4^{\mathrm T}+\mathsf g_2\zeta_{24}\mathsf f_4^{\mathrm T}
=\mathsf f_4\zeta_{41}\mathsf g_1^{\mathrm T}+\mathsf g_4\zeta_{41}\mathsf f_1^{\mathrm T}.\label{3forms}
\end{gather}
\end{theorem}
\begin{proof}
Direct calculation.
\end{proof}

\begin{remark}
\label{r:o}
Formula~\eqref{3forms} shows that the \emph{orientation} of the triangle edges is relevant here: edges
$12$, $24$ and~$41$, with this order of their vertices, are oriented as they must be as parts of the
boundary of triangle~$124$, according to the classical al\-ge\-braic-topolo\-gical def\/inition:
\begin{gather*}
\partial(ijk)=jk-ik+ij=ij+jk+ki.
\end{gather*}
\end{remark}

\begin{theorem}
There exists a~unique complex Euclidian metric in the space~$12345$ satisfying the following conditions:
\begin{enumerate}\itemsep 0pt
 \item
any two spaces corresponding to triangles~$ijk$ with disjoint interiors in Fig.~{\rm \ref{f:2}} or~{\rm \ref{f:3}} are
mutually orthogonal,
\item
\label{mii}
the scalar product within a~space~$ijk$, where $i<j<k$, is given by any of the expressions in the second
line of~\eqref{3forms}, with obvious changes $1\mapsto i$, $2\mapsto j$, $4\mapsto k$.
\end{enumerate}
\end{theorem}

\begin{proof}
First, we take one pentagon triangulation, for instance, the leftmost in Figs.~\ref{f:2} and~\ref{f:3}, and
\emph{define} the scalar product within each of the three spaces corresponding to its triangles according
to~\ref{mii}, while setting~-- again by def\/inition~-- these three spaces orthogonal to each other.
Then, formulas~\eqref{fg},~\eqref{XYsol} and Theorem~\ref{th:124} (with proper changes of indices) show
that the desired properties are preserved under any f\/lip in Fig.~\ref{f:2} or~\ref{f:3}.
\end{proof}

In particular, we see that, in addition to Conditions~\ref{i} and~\ref{ii}, the following condition holds.
\begin{cnd}
\label{iii}
The metric in a~space~$ijk$ depends only on $\zeta_i$, $\zeta_j$ and~$\zeta_k$.
\end{cnd}

We now turn to Condition~\ref{ii} and explain how to make the bases orthonormal within any space~$ijk$,
taking again $124$ as example.
We consider the \emph{Gramian}~-- matrix of scalar products~-- for the rows of the f\/irst matrix in~\eqref{f124}.
This Gramian is, according to~\eqref{3forms},
\begin{gather}
\label{G124}
G_{124}=(\mathbf f_{124},\mathbf f_{124})=\zeta_{12}\zeta_{42}^{-1}\zeta_{14}+\zeta_{14}\zeta_{42}^{-1}
\zeta_{12}=2\zeta_{12}\zeta_{42}^{-1}\zeta_{14},
\end{gather}
here the last equality is an easy exercise.
\begin{theorem}
There exists an $n\times n$ matrix $c_{124}$ such that
\begin{gather}
\label{ccT}
G_{124}=c_{124}c_{124}^{\mathrm T}.
\end{gather}
\end{theorem}
\begin{proof}
Gramian is, of course, symmetric: $G_{124}=G_{124}^{\mathrm T}$.
So, there are many ways to f\/ind such matrix~$c_{124}$ (and there are many such matrices for $n>1$).
For instance, one can use the \emph{Takagi decomposition}~\cite{Takagi24,Takagi25}: for a~complex symmetric
$n\times n$ matrix $A$ there exists a~unitary mat\-rix~$Q$ such that
\begin{gather*}
A=Q\Sigma Q^{\mathrm T},
\end{gather*}
where $\Sigma=\diag(\sigma_1,\dots,\sigma_n)$ with real nonnegative $\sigma_i$.
Or, otherwise, the reader can show, as an exercise using the theory of equivalence and normal forms of
matrices from textbook~\cite{Gantmacher}, that our \emph{nondegenerate} symmetric matrix~$G_{124}$ can even
be represented as a~squared, and also symmetric, matrix~$c_{124}$.
\end{proof}

Now the rows of matrix
\begin{gather}
\label{e124}
\mathbf e_{124}=c_{124}^{-1}\mathbf f_{124}
\end{gather}
give the desired orthonormal basis in~$124$; similarly in other spaces~$ijk$.
We get this way a~pentagon relation for orthogonal matrices, which we formulate as the following theorem.
\begin{theorem}
Matrices $P',\dots,T'$ of the following transitions between bases:
\begin{gather*}
P'
\begin{pmatrix}
\mathbf e_{124}
\\
\mathbf e_{234}
\end{pmatrix}
=
\begin{pmatrix}
\mathbf e_{123}
\\
\mathbf e_{134}
\end{pmatrix}
,
\qquad
\dots,
\qquad
T'
\begin{pmatrix}
\mathbf e_{125}
\\
\mathbf e_{235}
\end{pmatrix}
=
\begin{pmatrix}
\mathbf e_{123}
\\
\mathbf e_{135}
\end{pmatrix}
,
\end{gather*}
and identified with larger matrices so that they acquire structure~\eqref{bd}, are orthogonal and satisfy
the pentagon relation
$Q'P'=T'S'R'$.
Here bases~$\mathbf e_{ijk}$ are defined according to formulas of type~\eqref{e124}, and $c_{ijk}$ are
any matrices satisfying conditions of type~\eqref{ccT}.
\end{theorem}

Explicit formulas for $c_{ijk}$~-- and thus for $P',\dots,T'$~-- are not so easy to obtain.
Moreover, it may be more convenient, at least for an even~$n$ and if we want to pass to Grassmann algebras
in the style of Section~\ref{s:G} below, not to use orthonormal bases in spaces~$ijk$, but bases consisting
of \emph{isotropic} vectors~-- vectors of length zero.
This is what we are going to do, for $n=2$, in Section~\ref{s:2iso}.

\section[Explicit formulas for $n=2$ and isotropic bases]{Explicit formulas for $\boldsymbol{n=2}$ and isotropic bases} \label{s:2iso}

Let all $\zeta_i$ be symmetric $2\times2$ matrices, and consider again the basis~$\mathbf f_{124}$ in the
space~$124$ given by the f\/irst formula in~\eqref{f124} and the Gramian~\eqref{G124}.
For reasons that will be seen in Section~\ref{s:G}, we want to construct a~new basis
\begin{gather*}
\mathbf g_{124}=
\begin{pmatrix}
\mathbf g_{124}^{(1)}
\\
\mathbf g_{124}^{(2)}
\end{pmatrix}
=a_{124}\mathbf f_{124}
\end{gather*}
such that
\begin{gather}
\label{g124g124}
(\mathbf g_{124},\mathbf g_{124})=
\begin{pmatrix}
0&1
\\
1&0
\end{pmatrix}
.
\end{gather}
Condition~\eqref{g124g124} means, in particular, that both $\mathbf g_{124}^{(1)}$ and~$\mathbf
g_{124}^{(2)}$ have zero Euclidian lengths, i.e., are isotropic.

Denote the entries of~$a_{124}$ and~$G_{124}$ as
\begin{gather*}
a_{124}=
\begin{pmatrix}
x&y
\\
x'&y'
\end{pmatrix}
,
\qquad
G_{124}=
\begin{pmatrix}
\alpha&\beta
\\
\beta&\gamma
\end{pmatrix}
,
\end{gather*}
then~\eqref{g124g124} means the conditions
\begin{gather}
\label{a}
\begin{pmatrix}
x&y
\end{pmatrix}
G_{124}
\begin{pmatrix}
x
\\
y
\end{pmatrix}
=
\begin{pmatrix}
x'&y'
\end{pmatrix}
G_{124}
\begin{pmatrix}
x'
\\
y'
\end{pmatrix}
=0,
\qquad
\begin{pmatrix}
x&y
\end{pmatrix}
G_{124}
\begin{pmatrix}
x'
\\
y'
\end{pmatrix}
=1.
\end{gather}
The solution to~\eqref{a} is
\begin{gather}
\label{xy}
\begin{pmatrix}
x&y
\end{pmatrix}
=c
\begin{pmatrix}
-\beta{+}\sqrt{\beta^2{-}\alpha\gamma}&\alpha
\end{pmatrix}
,
\qquad
\begin{pmatrix}
x'&y'
\end{pmatrix}
=c'
\begin{pmatrix}
-\beta{-}\sqrt{\beta^2{-}\alpha\gamma}&\alpha
\end{pmatrix}
,
\end{gather}
with any $c$ and~$c'$ such that
\begin{gather}
\label{cc'}
cc'=\frac{1}{2\alpha(\alpha\gamma-\beta^2)}.
\end{gather}

Next, we def\/ine, the same way, isotropic bases in all other spaces corresponding to our triangles~$ijk$,
and def\/ine matrices $\tilde P,\dots,\tilde T$ by conditions
\begin{gather*}
\tilde P
\begin{pmatrix}
\mathbf g_{124}
\\
\mathbf g_{234}
\end{pmatrix}
=
\begin{pmatrix}
\mathbf g_{123}
\\
\mathbf g_{134}
\end{pmatrix}
,
\qquad
\dots,
\qquad
\tilde T
\begin{pmatrix}
\mathbf g_{125}
\\
\mathbf g_{235}
\end{pmatrix}
=
\begin{pmatrix}
\mathbf g_{123}
\\
\mathbf g_{135}
\end{pmatrix}
.
\end{gather*}
Due to the orthogonality conditions of the form~\eqref{perp} and isotropy and normalization conditions of
the form~\eqref{g124g124}, the matrices $\tilde P,\dots,\tilde T$ satisfy
\begin{gather*}
\tilde P^{\mathrm T}J\tilde P=\dots=\tilde T^{\mathrm T}J\tilde T=J
\end{gather*}
with
\begin{gather}
\label{I}
J=
\begin{pmatrix}
0&1&0&0
\\
1&0&0&0
\\
0&0&0&1
\\
0&0&1&0
\end{pmatrix}
.
\end{gather}
After identifying these matrices with $6\times6$ matrices, so that they acquire the form~\eqref{bd}, they
satisfy the pentagon equation
\begin{gather*}
\tilde Q\tilde P=\tilde T\tilde S\tilde R.
\end{gather*}
It remains to say that, for instance, $\tilde P$ is expressed the following way
\begin{gather}
\label{tP}
\tilde P=
\begin{pmatrix}
a_{123}&0
\\
0&a_{134}
\end{pmatrix}
P
\begin{pmatrix}
a_{124}^{-1}&0
\\
0&a_{234}^{-1}
\end{pmatrix}
,
\end{gather}
in terms of $P$ given by formula~\eqref{P}.
\begin{remark}
The freedom in this construction is, f\/irst, in choosing the sign of square roots in~\eqref{xy} and,
second, in choosing two values $c$ and~$c'$, obeying one condition~\eqref{cc'}, for every space~$ijk$.
\end{remark}

\section[Grassmann-Berezin calculus and pentagon relation in a~Grassmann algebra]{Grassmann--Berezin calculus and pentagon relation\\ in a~Grassmann algebra}
\label{s:GB}

A \emph{Grassmann algebra} over the f\/ield~$\mathbb C$ of complex numbers is an associative $\mathbb
C$-algebra with unity, generators~$x_i$ and relations
\begin{gather*}
x_i x_j=-x_j x_i.
\end{gather*}
In particular, $x_i^2=0$, so an element of a~Grassmann algebra is a~polynomial of degree $\le1$ in
each~$x_i$.
An \emph{even $($odd$)$ element} in Grassmann algebra consists, by def\/inition, only of monomials of even
(resp.\ odd) total degrees.

The \emph{exponent} is def\/ined by the standard Taylor series.
For example,
\begin{gather*}
\exp(x_1x_2+x_3x_4)=1+x_1x_2+x_3x_4+x_1x_2x_3x_4.
\end{gather*}

For every Grassmann generator~$x_i$, there are two derivations: \emph{left derivative}
$\frac{\partial}{\partial x_i}$ and \emph{right derivative}~$\frac{\overleftarrow{\partial}}{\partial
x_i}$.
These are $\mathbb C$-linear operations in Grassmann algebra def\/ined as
\begin{gather*}
\dfrac{\partial}{\partial x_i}(x_if)=f,
\qquad
(fx_i)\dfrac{\overleftarrow{\partial}}{\partial x_i}=f,
\end{gather*}
where the element~$f$ does not contain~$x_i$; in this case also, of course,
\begin{gather*}
\dfrac{\partial}{\partial x_i}f=f\dfrac{\overleftarrow{\partial}}{\partial x_i}=0.
\end{gather*}
More generally, there are the following Leibniz rules: for an even or odd~$f$,{\samepage
\begin{gather*}
\dfrac{\partial}{\partial x_i}(fg)=\dfrac{\partial}{\partial x_i}f\cdot g\pm f\dfrac{\partial}{\partial x_i}
g,
\qquad
(gf)\dfrac{\overleftarrow{\partial}}{\partial x_i}=g\cdot f\dfrac{\overleftarrow{\partial}}{\partial x_i}
\pm g\dfrac{\overleftarrow{\partial}}{\partial x_i}f,
\end{gather*}
namely, the plus is taken for an even~$f$, and the minus for an odd~$f$.}

In some situations, the derivation in Grassmann algebra constitutes an analogue not to usual
dif\/ferentiation, but to \emph{integration}.
In such cases, it is called \emph{Berezin integral}~\cite{B}; traditionally, Berezin integral is def\/ined
as the \emph{right} derivative.
In order to introduce the relevant notations, we also give its independent def\/inition: it is a~$\mathbb
C$-linear operator in Grassmann algebra satisfying
\begin{gather*}
\int\mathrm dx_i=0,
\qquad
\int x_i\,\mathrm dx_i=1,
\qquad
\int gh\,\mathrm dx_i=g\int h\,\mathrm dx_i,
\end{gather*}
if $g$ does not contain~$x_i$; multiple integral is understood as iterated one, where the Fubini rule is
applied according to the following model:
\begin{gather*}
\iint xy\,\mathrm dy\,\mathrm dx=\int x\left(\int y\,\mathrm dy\right)\mathrm dx=1.
\end{gather*}

The Grassmann-algebraic pentagon relation we are dealing with in this paper looks as follows:
\begin{gather}
\label{pGB}
\int\mathcal W_{1234}\mathcal W_{1345}\,\mathrm dx_{134}=\mathrm{const}\iiint\mathcal W_{1245}
\mathcal W_{2345}\mathcal W_{1235}\,\mathrm dx_{125}\,\mathrm dx_{235}\,\mathrm dx_{245}.
\end{gather}
Here Grassmann variables~$x_{ijk}$ (generators of the Grassmann algebra) are attached to all
two-faces~$ijk$ in Fig.~\ref{f:23}; the \emph{Grassmann weight}~$\mathcal W_{ijkl}$ of a~tetrahedron~$ijkl$
depends on (i.e., contains) the variables on its faces, e.g., $\mathcal W_{1234}$ depends on $x_{124}$,
$x_{234}$, $x_{123}$ and~$x_{234}$.
The integration in both sides of~\eqref{pGB} goes over variables on \emph{inner} faces (where the
tetrahedra are glued together), while the result depends on the variables on boundary faces.
Usually (compare, for instance,~\cite[formula~(30)]{11S}), there appears also a~numeric factor that cannot
be included in a~natural way in any~$\mathcal W$; we denote it~$\mathrm{const}$ in~\eqref{pGB}.

\section[Orthogonal operators from Grassmann-Gaussian exponents]{Orthogonal operators from Grassmann--Gaussian exponents}
\label{s:G}

We denote Grassmann variables~-- generators of a~Grassmann algebra~-- by letters $x$ and~$y$ with
subscripts.
The operators of left multiplication by~$x_i$:
\begin{gather}
\label{*x}
f\mapsto x_i f
\end{gather}
and left dif\/ferentiation w.r.t.~$x_i$:
\begin{gather}
\label{d/dx}
f\mapsto\frac{\partial}{\partial x_i}f
\end{gather}
satisfy fermionic anticommutation relations
\begin{gather*}
[x_i,\,x_j]_+=0,
\qquad
\left[\frac{\partial}{\partial x_i}, \frac{\partial}{\partial x_j}\right]_+=0,
\qquad
\left[x_i, \frac{\partial}{\partial x_j}\right]_+=\delta_i^j,
\end{gather*}
where the anticommutator of operators $A$ and~$B$ is
\begin{gather*}
[A,B]_+\stackrel{\rm def}{=}AB+BA.
\end{gather*}
Operators~\eqref{*x} and~\eqref{d/dx} make thus a~realization of \emph{fermionic creation-annihi\-la\-tion
operators}, and these operators generate a~\emph{Clifford algebra}, see~\cite[Chapter~I]{B}.

The \emph{canonical transformations} of the mentioned operators~-- linear transformations that preserve
the anticommutation relations~-- are given by orthogonal linear operators~\cite[Chapter~II]{B}.
Specif\/ically, right now we are going to deal with algebras generated by two variables.
For one algebra, called~$\mathfrak X$, the generators are called $x_1$ and~$x_2$, and the anticommutation
relations for the abovementioned operators can be written in the following matrix form
\begin{gather*}
\left[
\begin{pmatrix}
\frac{\partial}{\partial x_1}
\\
x_1
\\
\frac{\partial}{\partial x_2}
\\
x_2
\end{pmatrix}
, \,
\begin{pmatrix}
\frac{\partial}{\partial x_1}&x_1&\frac{\partial}{\partial x_2}&x_2
\end{pmatrix}
\right]_+=J,
\end{gather*}
where $J$ is our old matrix~\eqref{I}.
The operators $\frac{\partial}{\partial x_1}$, $x_1$, $\frac{\partial}{\partial x_2}$ and~$x_2$ are, of
course, linearly independent, so a~$4\times4$ matrix~$M$ of an endomorphism of the subspace spanned by
$\frac{\partial}{\partial x_1}$,~$x_1$,~$\frac{\partial}{\partial x_2}$,~$x_2$, relative to this basis, preserves
the anticommutation relations provided
\begin{gather*}
M^{\mathrm T}JM=J.
\end{gather*}
We write such canonical transformation as
\begin{gather*}
\begin{pmatrix}
\frac{\partial}{\partial x_1}
\\
x_1
\\
\frac{\partial}{\partial x_2}
\\
x_2
\end{pmatrix}
\mapsto M
\begin{pmatrix}
\frac{\partial}{\partial x_1}
\\
x_1
\\
\frac{\partial}{\partial x_2}
\\
x_2
\end{pmatrix}
.
\end{gather*}

One more algebra, called~$\mathfrak Y$, is a~copy of~$\mathfrak X$, with generators called $y_1$ and~$y_2$.
Note that we introduce below $\mathbb C$-linear operators $\mathcal A$, $\mathcal B$ and~$\mathcal C$
acting in the following spaces:
\begin{gather*}
\mathcal A\colon \ \mathfrak X\to\mathfrak X,
\qquad
\mathcal B\colon \ \mathfrak X\to\mathfrak Y,
\qquad
\mathcal C\colon \ \mathfrak Y\to\mathfrak Y,
\end{gather*}
while $A$, $B$ and~$C$ will be simply $4\times4$ matrices.

We are now interested in canonical transformations obtained from Gaussian exponents.
First, consider the operator~$\mathcal A$ of multiplying by the following Gaussian exponent, where $a$ is
a~numeric coef\/f\/icient:
\begin{gather}
\label{expa}
\mathcal A\colon \  f\mapsto\exp(ax_1x_2)f.
\end{gather}
As this operator obviously commutes with left multiplications by $x_1$ and~$x_2$, while
\begin{gather*}
\frac{\partial}{\partial x_1}\bigl(\exp(ax_1x_2)f\bigr)=\exp(ax_1x_2)\left(\frac{\partial}{\partial x_1}
+ax_2\right)f
\end{gather*}
and
\begin{gather*}
\frac{\partial}{\partial x_2}\bigl(\exp(ax_1x_2)f\bigr)=\exp(ax_1x_2)\left(-ax_1+\frac{\partial}
{\partial x_2}\right)f,
\end{gather*}
the following canonical transformation arises from pushing~\eqref{*x} and~\eqref{d/dx} from left to right
through~\eqref{expa}:
\begin{gather}
\label{A}
\begin{pmatrix}
\frac{\partial}{\partial x_1}
\\
x_1
\\
\frac{\partial}{\partial x_2}
\\
x_2
\end{pmatrix}
\mapsto\mathcal A^{-1}
\begin{pmatrix}
\frac{\partial}{\partial x_1}
\\
x_1
\\
\frac{\partial}{\partial x_2}
\\
x_2
\end{pmatrix}
\mathcal A=A
\begin{pmatrix}
\frac{\partial}{\partial x_1}
\\
x_1
\\
\frac{\partial}{\partial x_2}
\\
x_2
\end{pmatrix}
,
\qquad
\text{where}
\qquad
A=
\begin{pmatrix}
1&0&0&a
\\
0&1&0&0
\\
0&-a&1&0
\\
0&0&0&1
\\
\end{pmatrix}
.
\end{gather}

Next, consider the operator~$\mathcal B$ given by the following Berezin integral:
\begin{gather}
\label{expb}
\mathcal B\colon \  f\mapsto\int\exp(b_{11}x_1y_1+b_{12}x_1y_2+b_{21}x_2y_1+b_{22}
x_2y_2)f\,\mathrm dx_1 \mathrm dx_2.
\end{gather}
Again, a~simple calculation shows that pushing through~\eqref{expb} from left to right gives
\begin{gather}
\label{B}
\begin{pmatrix}
\frac{\partial}{\partial y_1}
\\
y_1
\\
\frac{\partial}{\partial y_2}
\\
y_2
\end{pmatrix}
\mapsto\mathcal B^{-1}
\begin{pmatrix}
\frac{\partial}{\partial y_1}
\\
y_1
\\
\frac{\partial}{\partial y_2}
\\
y_2
\end{pmatrix}
\mathcal B=B
\begin{pmatrix}
\frac{\partial}{\partial x_1}
\\
x_1
\\
\frac{\partial}{\partial x_2}
\\
x_2
\end{pmatrix}
,
\end{gather}
where
\begin{gather*}
B=\frac{1}{\Delta}
\begin{pmatrix}
0&-b_{11}\Delta&0&-b_{21}\Delta
\\
-b_{22}&0&b_{12}&0
\\
0&-b_{12}\Delta&0&-b_{22}\Delta
\\
b_{21}&0&-b_{11}&0
\end{pmatrix}
,
\qquad
\Delta=b_{11}b_{22}-b_{12}b_{21}.
\end{gather*}

And our third operator~$\mathcal C$ will be again like~\eqref{expa}:
\begin{gather}
\label{expc}
\mathcal C\colon \  f\mapsto\exp(cy_1y_2)f.
\end{gather}
The analogue of~\eqref{A} reads
\begin{gather}
\label{C}
\begin{pmatrix}
\frac{\partial}{\partial y_1}
\\
y_1
\\
\frac{\partial}{\partial y_2}
\\
y_2
\end{pmatrix}
\mapsto\mathcal C^{-1}
\begin{pmatrix}
\frac{\partial}{\partial y_1}
\\
y_1
\\
\frac{\partial}{\partial y_2}
\\
y_2
\end{pmatrix}
\mathcal C=C
\begin{pmatrix}
\frac{\partial}{\partial y_1}
\\
y_1
\\
\frac{\partial}{\partial y_2}
\\
y_2
\end{pmatrix}
,
\qquad
\text{where}
\qquad
C=
\begin{pmatrix}
1&0&0&c
\\
0&1&0&0
\\
0&-c&1&0
\\
0&0&0&1
\\
\end{pmatrix}
.
\end{gather}

The product~$\mathcal{CBA}$ of operators~\eqref{expc},~\eqref{expb} and~\eqref{expa} corresponds to the
following Grassmann weight of tetrahedron~$1234$:
\begin{gather}
\label{W}
\mathcal W_{1234}=\exp(ax_1x_2+b_{11}x_1y_1+b_{12}x_1y_2+b_{21}x_2y_1+b_{22}x_2y_2+cy_1y_2),
\end{gather}
where we identify
\begin{gather*}
x_1=x_{124},
\qquad
x_2=x_{234},
\qquad
y_1=x_{123},
\qquad
y_2=x_{134},
\end{gather*}
in the following sense:
\begin{gather}
\label{CBA}
\mathcal{CBA}\colon \  f\mapsto\int\mathcal W_{1234}f\,\mathrm dx_1\,\mathrm dx_2.
\end{gather}
This operator makes, according to~\eqref{A},~\eqref{B} and~\eqref{C}, the following canonical
transformation:
\begin{gather*}
\begin{pmatrix}
\frac{\partial}{\partial y_1}
\\
y_1
\\
\frac{\partial}{\partial y_2}
\\
y_2
\end{pmatrix}
\mapsto(\mathcal{CBA})^{-1}
\begin{pmatrix}
\frac{\partial}{\partial y_1}
\\
y_1
\\
\frac{\partial}{\partial y_2}
\\
y_2
\end{pmatrix}
\mathcal{CBA}=CBA
\begin{pmatrix}
\frac{\partial}{\partial x_1}
\\
x_1
\\
\frac{\partial}{\partial x_2}
\\
x_2
\end{pmatrix}
.
\end{gather*}
We will need the explicit expression for~$CBA$, which we also denote~$W_{1234}$:
\begin{gather}
\label{tozheCBA}
W_{1234}=CBA=\frac{1}{\Delta}
\begin{pmatrix}
cb_{21}&b_{11}(ac-\Delta)&-cb_{11}&b_{21}(ac-\Delta)
\\
-b_{22}&-ab_{12}&b_{12}&-ab_{22}
\\
cb_{22}&b_{12}(ac-\Delta)&-cb_{12}&b_{22}(ac-\Delta)
\\
b_{21}&ab_{11}&-b_{11}&ab_{21}
\end{pmatrix}
.
\end{gather}

Both sides of~\eqref{pGB} can be regarded as \emph{kernels of integral operators} acting on functions of
\emph{three} Grassmann variables.
To be more exact, there are three ``input'' variables, corresponding to three triangles in the leftmost
pentagon in Fig.~\ref{f:2} or~\ref{f:3}, and three ``output'' variables, corresponding to the triangles in
the rightmost pentagon in any of those f\/igures.
The mentioned integral operators are products of two or three operators of the form~\eqref{CBA},
respectively, for the l.h.s.\ or r.h.s.\ of~\eqref{pGB}, and each operator in these products corresponds to
a~tetrahedron.
\begin{theorem}
Let Gaussian exponents of the form~\eqref{W} be taken for all Grassmann tetrahedron weights entering
in~\eqref{pGB} $($of course, with its own parameters $a,\dots,c$ for each tetrahedron$)$.
Then equality~\eqref{pGB} holds, with a~proper factor~$\mathrm{const}$, provided the five corresponding
matrices~\eqref{tozheCBA} obey the pentagon relation in direct sum:
\begin{gather}
\label{5W}
W_{1345}W_{1234}=W_{1245}W_{2345}W_{1235}.
\end{gather}
\end{theorem}

\begin{proof}
It is known from general theory (see again~\cite{B}) that if a~$\mathbb C$-linear operator~$\mathcal K$ on
Grassmann algebra corresponds to a~(linear) canonical transformation~$K$ in the manner like our~$\mathcal
A$,~$\mathcal B$ and~$\mathcal C$ corresponded to~$A$,~$B$ and~$C$, then $\mathcal K$ is determined by~$K$
up to a~scalar factor.
So,~\eqref{pGB} follows if we take for~$K$ f\/irst l.h.s., and then r.h.s.\ of~\eqref{5W}.
\end{proof}

\section{A special case related to deformed exotic torsion}
\label{s:exotic}

In this section, we consider the case where all $\zeta_i$'s have the form
\begin{gather}
\label{lm}
\zeta_i=
\begin{pmatrix}
\lambda_i+\mu_i&\mathrm i\mu_i
\\
\mathrm i\mu_i&\lambda_i-\mu_i
\end{pmatrix}
,
\qquad
\mathrm i=\sqrt{-1}.
\end{gather}
\begin{remark}
All matrices of the form~\eqref{lm} commute.
The usual Jordan form of matrix~\eqref{lm}, if $\mu_i\ne0$, is $
\begin{pmatrix}
\lambda_i & 1
\\
0 & \lambda_i
\end{pmatrix}
$, see~\cite{Gantmacher}.
\end{remark}

Matrix~$G_{124}$ has, according to~\eqref{G124}, the same form~\eqref{lm}, namely
\begin{gather*}
G_{124}=
\begin{pmatrix}
\lambda+\mu&\mathrm i\mu
\\
\mathrm i\mu&\lambda-\mu
\end{pmatrix}
,
\qquad
\lambda=\frac{2\lambda_{21}\lambda_{41}}{\lambda_{42}},
\qquad
\mu=\frac{2(\lambda_{41}^2\mu_{21}-\lambda_{21}^2\mu_{41})}{\lambda_{42}^2},
\end{gather*}
where we write, in the style of formula~\eqref{zij},
\begin{gather*}
\lambda_{ij}=\lambda_i-\lambda_j,
\qquad
\mu_{ij}=\mu_i-\mu_j.
\end{gather*}
Possible matrices~$a_{124}$ are, according to Section~\ref{s:2iso}, as follows:
\begin{gather*}
a_{124}=\frac{1}{2\lambda}\diag\big(c_{124},c_{124}^{-1}\big)
\begin{pmatrix}
2&2\mathrm i
\\
\lambda-\mu&-\mathrm i(\lambda+\mu)
\end{pmatrix}
,
\end{gather*}
where $c_{124}$ is an arbitrary nonzero constant.

Then we calculate the matrices $a_{234}$, $a_{123}$ and~$a_{134}$ in the same way as we have done
for~$a_{124}$, only changing the indices as required, then the matrix~$P$ according to~\eqref{P}, and
f\/inally $\tilde P$ according to~\eqref{tP}.
The result is
\begin{gather*}
\tilde P=\diag\big(c_{124},c_{124}^{-1},c_{234},c_{234}^{-1}\big)\,\hat P \diag\big(c_{123}^{-1},c_{123}
,c_{134}^{-1},c_{134}\big),
\end{gather*}
where
\begin{gather}
\label{tPe'}
\hat P=
\begin{pmatrix}
\dfrac{\lambda_{32}\lambda_{41}}{\lambda_{31}\lambda_{42}}& 0& -\dfrac{\lambda_{32}}{\lambda_{31}}& 0
\vspace{2mm}\\
-\dfrac{\lambda_{21}A}{\lambda_{31}\lambda_{32}\lambda_{42}^2}& 1& -\dfrac{A}{\lambda_{31}\lambda_{42}
\lambda_{43}}& -\dfrac{\lambda_{21}\lambda_{43}}{\lambda_{32}\lambda_{42}}
\vspace{2mm} \\
\dfrac{\lambda_{21}\lambda_{43}}{\lambda_{31}\lambda_{42}}& 0& \dfrac{\lambda_{32}}{\lambda_{31}}
& 0
\vspace{2mm}\\
\dfrac{\lambda_{41}A}{\lambda_{31}\lambda_{42}^2\lambda_{43}}&1&-\dfrac{A}{\lambda_{31}
\lambda_{42}\lambda_{43}}&\dfrac{\lambda_{41}}{\lambda_{42}}
\end{pmatrix}
,
\end{gather}
and
\begin{gather*}
A=\left|
\begin{matrix}
1&\lambda_1&\lambda_1^2&\mu_1
\\
1&\lambda_2&\lambda_2^2&\mu_2
\\
1&\lambda_3&\lambda_3^2&\mu_3
\\
1&\lambda_4&\lambda_4^2&\mu_4
\end{matrix}
\right|
\end{gather*}
(again, Maxima 
has helped much in these calculations).

The zeros in matrix~\eqref{tPe'}, when compared with~\eqref{tozheCBA}, show that $\zeta_i$'s of the
form~\eqref{lm} correspond to the case where
\begin{gather}
\label{u}
ac-b_{11}b_{12}+b_{12}b_{21}=0.
\end{gather}
This \looseness=-1 is exactly the case for the Grassmann tetrahedron weights already introduced in~\cite{11S, 2deform}.
Recall that the starting point in these papers was Grassmann weights related to Reidemeister torsion of
some exotic chain complexes, and then a~deformation, like in~\cite[formula~(23)]{11S}, makes them (up to
a~numeric factor) a~Grassmann--Gaussian exponent for which the condition~\eqref{u} can be easily checked.
Actually, any Grassmann tetrahedron weight obeying~\eqref{u} can be obtained from weights
in~\cite{11S, 2deform} by a~proper scaling $x_{ijk}\mapsto\mathrm{const}_{ijk}x_{ijk}$ of Grassmann
generators.
This shows that the present paper is about further nontrivial deformations of the exotic torsion weights.

\subsection*{Acknowledgements}

The authors thank Rinat Kashaev for useful discussions, and the referees for careful reading of the f\/irst
version of the paper and valuable comments.

\pdfbookmark[1]{References}{ref}
\LastPageEnding

\end{document}